\documentclass[journal,10pt,twoside]{IEEEtran}

\usepackage{cite}
\usepackage{acro}

\usepackage{graphicx}
\graphicspath{../Figure/}
\DeclareGraphicsExtensions{.pdf,.jpeg,.png,.jpg, .svg}
\usepackage[dvipsnames]{xcolor}
\usepackage[font=footnotesize]{caption}
\usepackage{tikz}
\usepackage{pgfplots}
\pgfplotsset{compat=1.14} 
\usepgfplotslibrary{groupplots}
\usetikzlibrary{calc}

\usepackage{amsmath}
\usepackage{amsfonts}
\usepackage{bm}
\usepackage{mathrsfs}
\usepackage{mathtools}

\usepackage{array}
\usepackage{theoremref}

\usepackage{algorithmic}
\usepackage[algo2e]{algorithm2e}
\usepackage{algorithm}

\newlength\myindent
\newcommand\bindent{%
  \begingroup
  \setlength{\itemindent}{\myindent}
  \addtolength{\algorithmicindent}{\myindent}
}
\newcommand\eindent{\endgroup}
\newlength\myindentt
\newcommand{\INDSTATE}[1][1]{\STATE\hspace{#1\myindentt}}
\usepackage{array}
\usepackage{textcomp}
\usepackage{stfloats}

\usepackage{hyperref} 
\usepackage{url}
\usepackage{verbatim}

	\DeclareMathAlphabet\mathbfcal{OMS}{cmsy}{b}{n}
	\definecolor{myMagneta}{rgb}{1, 0, 1}
    
    \newcounter{lemma}

	\renewcommand{\b}[1]{\mathbf{#1}}
	\newcommand{\rvec}[1]{\mathbf{\underline{#1}}}
    \newcommand{\grvec}[1]{\bm{\underline{#1}}}
	\renewcommand{\vec}[1]{\underline{#1}}

     \newcommand{\highlight}[1]{#1}
    \newcommand{\I}{\mathrm{I}}
    
    \newcommand{\E}{\mathbb{E}}

    \newcommand\norm[1]{\left\lVert#1\right\rVert}
    \newcommand{\pauliTensor}{\overrightarrow{\sigma}}

    \newcommand{\Hb}{\mathbf{H}}
    
    \newcommand{\Jb}{\mathbf{J}}

    \newcommand{\Hbhat}{\mathbf{\hat{H}}}

    \newcommand{\Hbr}{\mathbf{H}^\mathrm{r}}
    
    \newcommand{\Rb}{\mathbf{R}}
    \newcommand{\Rbhat}{\mathbf{\hat{R}}}
    
    \newcommand{\Sb}{\mathbf{S}}
    \newcommand{\Sbhat}{\mathbf{\hat{S}}}
    
    \newcommand{\Sp}{\mathrm{S^p}}
    \renewcommand{\sp}{\vec{s}^\mathrm{p}}
    \newcommand{\p}{\mathrm{p}}
    
    \newcommand{\Deltap}{\Delta\mathrm{p}}
    \newcommand{\Deltaptot}{\Delta \p_{\text{tot}}}
    
    \newcommand{\Ks}{K_\mathrm{s}}
    \newcommand{\Kp}{K_\mathrm{p}}

    \newcommand{\Gammab}{\mathbf{\Gamma}}

    \newcommand{\Gb}{\mathbf{G}}
    \newcommand{\Gbhat}{\mathbf{\hat{G}}}

    \newcommand{\CN}{\mathcal{CN}}

    \renewcommand{\highlight}[1]{#1}
    
    \newcommand{\hlightRI}[1]{#1}
    \newcommand{\hlightRII}[1]{#1}

\DeclareAcronym{SOP}{
short=SOP,
long= state of polarization,
}

\DeclareAcronym{AIR}{
short=AIR,
long= achievable information rate,
}
\DeclareAcronym{GD}{
short=GD,
long= gradient descent,
}
\DeclareAcronym{EDFA}{
short=EDFA,
long=  erbium-doped fiber-amplifier,
}

\DeclareAcronym{WSS}{
short=WSS,
long=  wavelength selective switches,
}

\DeclareAcronym{DSP}{
short=DSP,
long= digital signal processing,
}

\DeclareAcronym{PMD}{
short=PMD,
long= polarization-mode dispersion,
}
\DeclareAcronym{PDL}{
short=PDL,
long= polarization-dependent loss,
}
\DeclareAcronym{DOF}{
short=DOF,
long= degrees of freedom,
}

\DeclareAcronym{CMA}{
short=CMA,
long= constant modulus algorithm,
}
\DeclareAcronym{MCMA}{
short= MCMA,
long= modified {CMA},
}
\DeclareAcronym{PM}{
short= PM,
long= polarization multiplexed,
}
\DeclareAcronym{LS-DDLMS}{
short= LS-DDLMS,
long= LS-DDLMS,
}
\DeclareAcronym{ML}{
short= ML,
long= maximum likelihood}
\DeclareAcronym{PDM}{
short= PDM,
long= polarization-division multiplexed,
}
\DeclareAcronym{SVD}{
short= SVD,
long= singular value decomposition}

\DeclareAcronym{MMA}{
short=MMA,
long= multi-modulus algorithm,
}
\DeclareAcronym{RDE}{
short=RDE,
long= radially directed equalizer,
}

\DeclareAcronym{QAM}{
short=QAM,
long= quadrature amplitude modulation,
}
\DeclareAcronym{PM-16-QAM}{
short=PM-$16$-QAM,
long= polarization-multiplexed $16$ quadrature amplitude modulation,
}
\DeclareAcronym{SER}{
short=SER,
long= symbol error rate,
}
\DeclareAcronym{PTC}{
short=PTC,
long= polarization-time code,
}
\DeclareAcronym{DD}{
short=DD,
long= decision-directed,
}
\DeclareAcronym{LMS}{
short=LMS,
long= least mean squares,
}
\DeclareAcronym{DDLMS}{
short=DDLMS,
long= decision-directed least mean squares,
}
\DeclareAcronym{DD-Kabsch}{
short=DD-{K}absch,
long=  decision-directed {K}absch,
}

\DeclareAcronym{DD-Czegledi}{
short=DD-{C}zegledi,
long=  decision-directed {C}zegledi,
}
\DeclareAcronym{SW-Kabsch}{
short= SW-Kabsch,
long= sliding window Kabsch,
}

\DeclareAcronym{SW-ULS}{
short= SW-ULS,
long= sliding window unitary least square error,
}

\DeclareAcronym{LS-SW-ULS}{
short= LS-SW-ULS,
long = LS-SW-ULS,
}

\DeclareAcronym{SW-LS}{
short= SW-LS,
long= sliding window least squares,
}
\DeclareAcronym{MMSE}{
short=MMSE,
long= minimum mean square error,
}
\DeclareAcronym{LS}{
short=LS,
long= least squares,
}

\DeclareAcronym{DP}{
short=DP,
long= dual-polarization,
}

\DeclareAcronym{DP-PDL}{
short=DP-PDL,
long= dual-polarization PDL,
}

\DeclareAcronym{PN}{
short=PN,
long= phase noise,
}

\DeclareAcronym{GN}{
short=GN,
long= Gaussian noise,
}

\DeclareAcronym{NLI}{
short=NLI,
long= nonlinear impairments,
}

\DeclareAcronym{MIMO}{
short=MIMO,
long=  multiple-input multiple-output,
}

\DeclareAcronym{SNR}{
short=SNR,
long= signal-to-noise ratio,
}

\DeclareAcronym{MI}{
short=MI,
long= mutual information,
}

\DeclareAcronym{pdf}{
short=pdf,
long= probability density function,
}

\DeclareAcronym{AWGN}{
short=AWGN,
long= additive white Gaussian noise,
}

\DeclareAcronym{ASE}{
short=ASE,
long= amplified spontaneous emission,
}

\DeclareAcronym{QPSK}{
short=QPSK,
long= quadrature phase-shift keying,
}
\begin{document}
\title{Polarization Tracking in the Presence of PDL and Fast Temporal Drift}
\author{Mohammad~Farsi,~\IEEEmembership{Student Member,~IEEE,} Christian Häger,~\IEEEmembership{Member,~IEEE}, Magnus~Karlsson,~\IEEEmembership{Senior Member,~IEEE}, and Erik~Agrell,~\IEEEmembership{Fellow,~IEEE.} 
\thanks{M. Farsi, C. Häger, and E. Agrell are with the Department
of Electrical Engineering, Chalmers University of Technology, SE-41296 Gothenburg, Sweden (e-mails: \{farsim, christian.haeger, agrell\}@chalmers.se)}
\thanks{M. Karlsson is with the Department of Microtechnology
and Nanoscience, Chalmers University of Technology, SE-41296 Gothenburg,
Sweden (e-mail: magnus.karlsson@chalmers.se).}
\thanks{This work was supported in part by the Knut and Alice Wallenberg Foundation under grant no. 2018.0090 and the Swedish Research Council under grants no. 2020-04718 and 2021-03709.}
\thanks{The computations were enabled by resources provided by the Swedish National Infrastructure for Computing (SNIC), partially funded by the Swedish Research Council through grant agreement no. 2018-05973}
}



\maketitle

\begin{abstract}
In this paper, we analyze the effectiveness of polarization tracking algorithms in optical transmission systems suffering from fast state of polarization (SOP) rotations and polarization-dependent loss (PDL). While most of the gradient descent (GD)-based algorithms in the literature may require step size adjustment when the channel condition changes, we propose tracking algorithms that can perform similarly or better without parameter tuning. Numerical simulation results show higher robustness of the proposed algorithms to SOP and PDL drift compared to GD-based algorithms, making them promising candidates to be used in aerial fiber links where the SOP can potentially drift rapidly, and therefore becomes challenging to track.
\end{abstract}

\section{Introduction}
\IEEEPARstart{A}{dvanced} \ac{DSP} enables the use of multilevel modulation formats and polarization-division multiplexed transmission to achieve high spectral efficiency \cite{Kikuchi_coherent_optic:2016}. 
However, spectrally-efficient systems have a lower tolerance to fiber impairments such as \ac{SOP} drift and \ac{PDL}, which need to be adaptively estimated and tracked at the receiver \cite{Roberts:2009}. 

In an optical link, the random \ac{SOP} drift originates from the random variation of environmental conditions (mechanical/thermal stress, weather conditions, splices, etc.) and internal imperfections (asymmetry of the core, manufacturing process errors, etc.) of the fiber. Experimental measurements show that the \ac{SOP} drift can be extremely slow (days and hours) in buried fibers \cite{Karlsson:2000} and very fast (microseconds) in aerial fibers \cite{Waddy:2001,Wuttke:2003, Krummrich:2016}. \hlightRI{For instance, field measurements of an aerial fiber link have revealed that the rotation rate of \ac{SOP} might get up to 5.1 Mrad/s due to lightning strikes \cite{Charlton:2017}.}\label{cmr1:sop_drift_speed} In the literature, the \ac{SOP} fluctuation is often modeled as randomly chosen rotations without drift \cite{kikuchi_sop_model:2008,roudas_sop_model_demult:2009} or cyclic/quasi-cyclic deterministic rotation \cite{heismann_sop_model:1995,Savory:2008,muga_sop_adaptive_polux:2014}. Recently, an experimentally verified model was proposed in \cite{czegledi_sop_drift_model:2016}, where \ac{SOP} drift is described by a random walk on the Poincar\'e sphere.

In a long-haul fiber, several \ac{PDL}-inducing components (e.g., isolators, amplifiers, multiplexers, and couplers) are in place, and the overall \ac{PDL} might aggregate to several dBs. Historically, \ac{PDL} has been modeled as a concatenation of many randomly oriented \ac{PDL} elements along with the fiber, and its statistics have been extensively studied in \cite{gisin_pdl_statistics:1995,mecozzi_pdl_stat:2002,galtarossa_pdl_statistics:2003,vinegoni_pdl_statistics:2004}, describing the aggregated \ac{PDL} with a Maxwellian distribution. The impact of the average \ac{PDL} on the performance of the optical link is studied in \cite{shtaif_pdl_performance_loss:2008,duthel_pdl_impact:2008}. The interplay of \ac{PDL} with polarization-mode dispersion and Kerr nonlinearity has been studied in \cite{shtaif_pdl_vs_pmd:2005}, and \cite{Rossi_pdl_vs_NL:2014}, respectively.

The \ac{SOP} drift accompanied by \ac{PDL} \hlightRII{not only} results in a time-varying power imbalance between the two polarizations, \label{cmr2:orthogonality}\hlightRII{ but also it causes optical \ac{SNR} fluctuations and breaks the orthogonality between the two polarizations \cite{Xie:2010}.} While it would be feasible to resolve the static \ac{SOP} and \ac{PDL}, the time-varying nature of \ac{SOP} and the aggregated \ac{PDL} makes the polarization tracking challenging at the receiver. 

The \ac{GD}-based algorithms has been widely applied in both wireless and optical fields for adaptive tracking, and the most popular one is the \ac{CMA} \cite{Godard:1980} and its variants known as \ac{MCMA} \cite{oh_MCMA:1995} and \ac{MMA} \cite{Yang_MMA:2002}. Although \ac{CMA} is immune to \ac{PN} and has low computational complexity, it is modulation dependent and suffers from phase ambiguity and the so-called singularity problem \cite{Kikuchi:2011}, where the equalizer converges only on one of two polarizations.
The \ac{DDLMS} algorithm~\cite{Lucky_DDLMS:1966,Savory:2010} removes the modulation format dependence of the blind algorithms \cite{Savory:2008}, and the phase ambiguity can be resolved by differential encoding/decoding at the expense of a performance degradation \cite[Sec. 2.6.1]{seimetz_diffdecoding:2009}.
Although many have tried to overcome the singularity problem of \ac{CMA} \cite{liu_cma_init:2009,vgenis_nonsingular_cma:2009}, its solution remains an open research problem. The pilot-aided hybrid algorithms~\cite{Morsy-Osman:2012,Faruk:2010} ensure reliable and fast convergence by proper tap initialization of the blind algorithms. 
For instance, in \cite{Faruk:2010}, a pilot-based filter tap initialization using the \ac{LMS} \cite{LMS_Widrow1971} algorithm has been proposed to address the singularity problem. \label{cmr1:Wiener}\hlightRI{For a linear time-invariant system with stationary noise, the \ac{LMS} solution converges to the well-known Wiener filter solution \cite{wiener:1949},\cite[Ch. 13.2]{proakis:2000}.} 
The memoryless dual-polarization channel in the presence of \ac{SOP} drift and a negligible amount of \ac{PDL} can be described by a time-varying complex unitary matrix. Most of the tracking algorithms in the literature (e.g., \ac{CMA}, \ac{MMA}, \ac{DDLMS}, etc.) are designed for a general complex channel matrix without constraining their channel estimation matrix to be unitary. The tracking performance can be improved by explicitly taking the unitary nature of the SOP rotations into account when designing tracking algorithms (even if the overall channel is not exactly unitary). For example, the block-wise \ac{DD-Kabsch} algorithm was proposed in \cite{Louchet:2014} to address the unitary constraints of the channel.
In \cite{Czegledi:2016}, a unitary \ac{GD}-based decision-directed joint \ac{PN} and \ac{SOP} tracking algorithm was proposed, where the authors have shown a higher polarization drift tolerance than the Kabsch and \ac{CMA}-based algorithms. We refer to the algorithm in \cite{Czegledi:2016} as \ac{DD-Czegledi}. 

This paper, which extends the conference paper \cite{Farsi:2022}, proposes two polarization tracking algorithms called \ac{SW-Kabsch} and \ac{SW-LS}, described in Sections \ref{subSec:slidingKabsch} and \ref{subSec:SW-LS}, respectively.
When the \ac{PDL} is negligible, \ac{SW-Kabsch} shows higher \ac{SOP} drift tolerance than \ac{SW-LS} and the \ac{GD}-based algorithms (e.g., \ac{DD-Czegledi}, \ac{CMA} and its variants). However, considering \ac{PDL}, the channel is no longer unitary, and \ac{SW-LS} shows the highest \ac{SOP} drift tolerance. Moreover, unlike \ac{GD}-based algorithms that require step size adjustment, the proposed algorithms require no further parameter tuning, making them potential candidates for memoryless fast drifting optical systems.

Although the literature is replete with time-invariant \ac{PDL} models \cite{mecozzi_pdl_stat:2002,galtarossa_pdl_statistics:2003}, the time-varying nature of \ac{PDL} is not well-studied. Inspired by the previous work \cite{czegledi_sop_drift_model:2016}, we introduce a channel model called \ac{DP-PDL} (described in Section~\ref{sec:system_model}) that accounts for memoryless time-varying \ac{PDL}.

\textit{Notation:} Column vectors are denoted by underlined letters $\vec{x}$ and matrices by uppercase roman letters $\mathrm{X}$. We use bold-face letters $\mathbf{x}$ for random quantities and the corresponding nonbold letters $x$ for their realizations.
Sets are denoted by uppercase calligraphic letters $\mathcal{X}$. The Frobenius norm is denoted by $\norm{\cdot}$ and the expectation over random variables is denoted by $\E[\cdot]$. The real zero-mean multivariate Gaussian distribution is denoted by $\rvec{x}\sim\mathcal{N}(\vec{0},\Lambda_{\vec{x}})$ and the complex zero-mean circularly symmetric Gaussian distribution of a vector is denoted by $\rvec{x}\sim\CN(\vec{0},\Lambda_{\vec{x}})$, where $\Lambda_{\vec{x}}$ is the covariance matrix. \begin{figure}
    \centering
    \includegraphics[scale=.26]{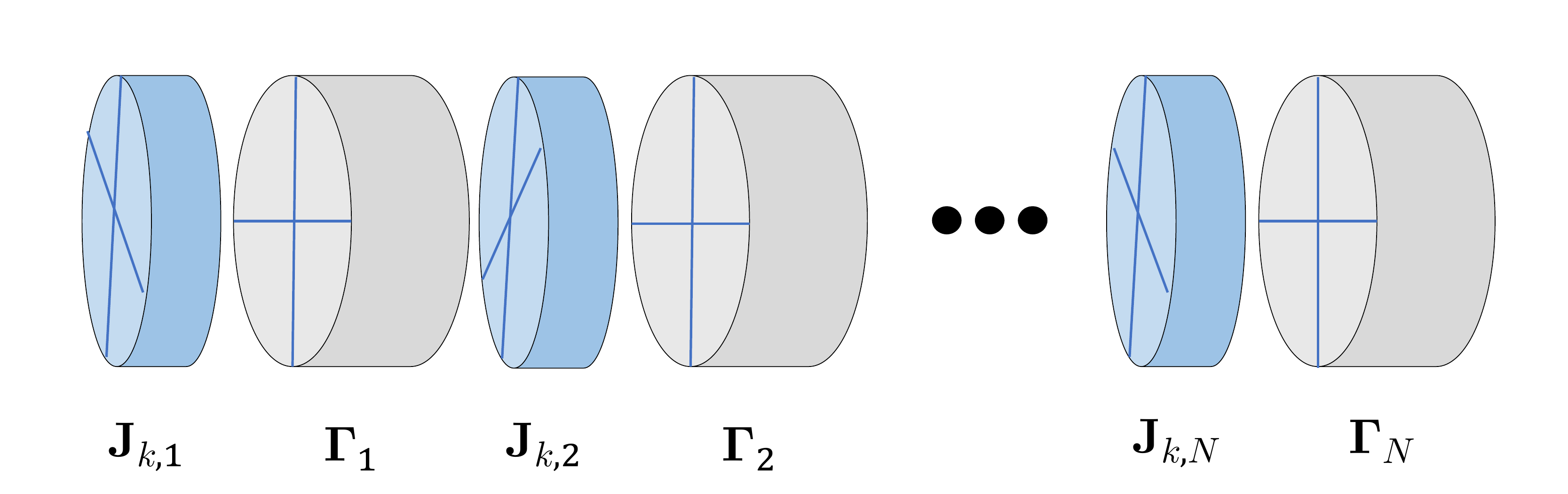}
    \caption{The \ac{DP-PDL} model of a long fiber at time $k$ as a concatenation of $N$, each containing a \ac{PDL} component $\Gammab_{n}$ coupled with an \ac{SOP} drift element $\Jb_{k,n}$.}
    \label{fig:PhysicalModel}
\end{figure}
\section{System Model}\label{sec:system_model}
We consider dual-polarization transmission in the presence of \ac{PDL}, \ac{SOP} drift, and \ac{ASE} noise at the receiver. We also assume that nonlinearities are negligible and chromatic dispersion is compensated. We particularly study the memoryless dual-polarization channel where polarization-mode dispersion is not considered. 
To model the channel, we take the experimentally verified \ac{SOP} drift model in \cite{czegledi_sop_drift_model:2016} and combine it with concatenated \ac{PDL} elements. \label{cmr1:deltap_part1}\hlightRI{This model can be regarded as a spherical analogy of a two-dimensional Wiener process.}

The transmitted signal at time $k$ is a $2$-dimensional random vector $\rvec{s}_k$, which takes on values from a set $\mathcal{S}=\{\vec{c}_1,\vec{c}_2,...,\vec{c}_M\}$ of complex, zero-mean, equiprobable constellation points. 
The vector of received complex samples $\rvec{x}_k$ can be expressed as
\begin{equation}\label{eq:System_Model}
	\rvec{x}_k=\Hb_k\rvec{s}_k+\rvec{z}_k,
\end{equation}
where $\Hb_k$ is a $2\times 2$ complex channel matrix, and $\rvec{z}_k \sim \mathcal{CN}(0,\sigma_z^2 \I_2)$. The physical model at time $k$ is shown in Fig.~\ref{fig:PhysicalModel}, where a link with $N$ segments, each consisting of a \ac{SOP} element $\Jb_{k,n}$ and a \ac{PDL} element $\Gammab_{n}$, where $n$ is the segment index. The \ac{DP-PDL} channel matrix can be described by a $2\times 2$ complex-valued matrix as 
\begin{align}\label{eq:channel_model}
\Hb_k = \Gammab_{N}\Jb_{k,N}\cdots\Gammab_{1}\Jb_{k,1} = \prod_{n=1}^{N} \Gammab_{n}\Jb_{k,n},
\end{align}
where $\Gammab_{n}$ is a $2\times 2$ positive real-valued diagonal matrix modeling the power imbalance induced by \ac{PDL}. For the sake of simplicity, we assume that a time-invariant deterministic matrix can describe each \ac{PDL} component as \cite{mecozzi_pdl_stat:2002} 
\begin{align}
    \Gammab_{n} = \begin{bmatrix}
    \sqrt{1+\gamma_n}& 0\\
    0 &\sqrt{1-\gamma_n}\\
    \end{bmatrix}
\end{align}
where $0\le \gamma_n\le 1$ is each segment's \ac{PDL} ratio indicating that in the extreme case, only one active polarization will remain ($\gamma_n = 1$). Moreover, $\Jb_{k,n}$ is a random $2\times 2$ unitary matrix accounting for \ac{SOP} drift defined in \cite{czegledi_sop_drift_model:2016} as
\begin{align}\label{eq:j_update}
    \Jb_{k+1,n}= \exp(-j\grvec{\alpha}_{k,n}\cdot\pauliTensor)\Jb_{k,n}, 
\end{align}
where $\exp{(\cdot)}$ is the matrix exponential and 
\begin{align}\label{eq:innovationVector}
    \grvec{\alpha}_{k,n} \sim \mathcal{N}(\vec{0},\sigma_\p^2 \I_3),
\end{align}
where $\sigma_\p^2 = 2\pi\Deltap T $ and $\Deltap$ is referred to as the \textit{polarization linewidth} determining the speed of the \ac{SOP} drift and $T$ is the symbol duration. Finally, $\pauliTensor = (\sigma_1,\sigma_2,\sigma_3)$ is a tensor of the Pauli spin matrices \cite[eq. (2.5.19)]{damaskBook:2005} 
\begin{align}\label{eq:pauliMatrices}
    \sigma_1 = \begin{bmatrix}
    1 & 0\\
    0 & -1
    \end{bmatrix},
    \quad
    \sigma_2 = \begin{bmatrix}
    0 & 1\\
    1 & 0
    \end{bmatrix},
    \quad 
    \sigma_3 = \begin{bmatrix}
    0 & -j\\
    j & 0
    \end{bmatrix}.
\end{align}
The total polarization linewidth scales with $N$ and can be defined as 
\begin{align}
    \Deltaptot = N\cdot\Deltap.
\end{align}

\begin{figure*}
	\centering
         \includegraphics{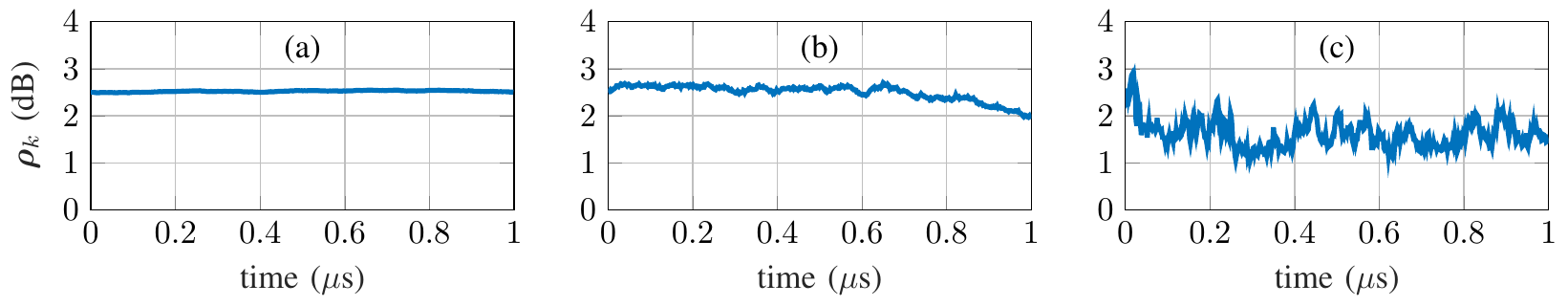}
	\caption{\label{fig:PDL_Evolution} The evolution of aggregated \ac{PDL} ratio $\boldsymbol{\rho}_k$ in dB for different \ac{SOP} drifts in a time period of $1$ microsecond is plotted. The symbol rate is $28$ Gbaud (i.e., the symbol duration $T = 3.57 \cdot 10^{-11}$) and $N=20$. (a) $\Deltaptot\cdot T = 10^{-8}$, (b) $\Deltaptot\cdot T = 3.57\cdot 10^{-6}$, and (c) $\Deltaptot\cdot T = 3.57\cdot 10^{-5}$. }
\end{figure*}

For the \ac{DP-PDL} channel, we define the segment-wise \ac{PDL} in dB as
\begin{align}\label{eq:varphi}
    {\varphi}_n = 10\log_{10}\left(\frac{1+\gamma_n}{1-\gamma_n}\right),
\end{align}
and the aggregated \ac{PDL} ratio at time $k$ as 
\begin{align}\label{eq:agg_PDL}
\boldsymbol{\rho}_k = \frac{\norm{\boldsymbol{\lambda}_k^{\max}}^2}{\norm{\boldsymbol{\lambda}_k^{\min}}^2},
\end{align}
where $\boldsymbol{\lambda}_k^{\max}$ and $\boldsymbol{\lambda}_k^{\min}$ are the \label{cmr2:cm2_singularity}\hlightRII{singular values} of $\Hb_{k}$. The average aggregated \ac{PDL} over $K$ transmitted symbols in dB is defined as
\begin{align}
\bar{\rho} = 10\log_{10}\left(\E_{\Hb}\left[\frac{1}{K}\sum_{k=0}^{K-1} \boldsymbol{\rho}_k\right]\right).
\end{align}
Although each \ac{PDL} component of the fiber is assumed to be constant, the aggregated \ac{PDL} $\boldsymbol{\rho}_k$ of the channel will drift with time.
Fig. \ref{fig:PDL_Evolution} illustrates the dependence of the aggregated \ac{PDL} dependence on the \ac{SOP} drift. The \ac{PDL} evolution is shown in a window of one microsecond with three different \ac{SOP} drift speeds showing that $\boldsymbol{\rho}_k$ is strongly dependent on the speed of the \ac{SOP} drift making the channel tracking challenging.

\section{Proposed Algorithms}\label{sec:Tracking_Algorithms}
Since the received data suffers from a time-varying \ac{SOP} drift, adaptive channel tracking is needed to recover the transmitted data. This section describes the proposed channel tracking algorithms. First, an adaptive unitary tracking algorithm is proposed, which performs best when \ac{PDL} is negligible (e.g., $\Gammab_{n} = \I_2\quad \forall{n}$); after that, a data-aided estimation algorithm is proposed that estimates and tracks both \ac{SOP} and \ac{PDL} of the \ac{DP-PDL} channel.

\subsection{Decision-directed Sliding Window Kabsch Algorithm} \label{subSec:slidingKabsch}
In this section, we introduce a modification to the \ac{DD-Kabsch} algorithm \cite{Louchet:2014} and make it suitable for polarization tracking in fast drifting channels. The key idea is to add a sliding window called \textit{equalization window} with a size of $L$ and update the channel estimate for each window.
We define
\begin{align}\label{eq:rxSym_Def}
    \b{X}_{k} &= [\rvec{x}_k,\rvec{x}_{k+1},\dots,\rvec{x}_{k+L-1}],\nonumber\\
    \Sbhat_{k} &= [\rvec{\hat{s}}_k,\rvec{\hat{s}}_{k+1},\dots,\rvec{\hat{s}}_{k+L-1}],
\end{align}
where $\b{X}_k$ is a $2\times L$ matrix of received symbols and $\Sbhat_k$ is a $2\times L$ matrix of estimated received symbols.
Ignoring the \ac{PDL}, the channel matrix can be expressed as
\begin{align}
    \Hb_{k+1} = \b{R}_k\Hb_{k},
\end{align}
where
\begin{align}
    \Rb_{k} = \highlight{\Jb_{k,N}\Jb_{k,N-1}\cdots\Jb_{k,1}},
\end{align}
is obtained from \eqref{eq:channel_model} by substituting~${\Gammab_{n}=\I_2}$.

Although the channel $\Hb_k$ changes for every $k$, a constant estimated channel $\Hbhat_k$ is used over an equalization window of length $L$ to obtain the required decision-directed symbols for the tracking algorithm. This can be justified when the speed of the drift is not extremely high, and $L$ is small. Then the transmitted symbols in an equalization window can be estimated by the previous estimate of the channel $\Hbhat_{k}$ and using the minimum Euclidean distance criterion
\begin{align}\label{eq:dd_update}
     \hat{\rvec{s}}_{k} = \arg\min\limits_{\vec{c}\in\mathcal{S}}\norm{\Hbhat^{-1}_{k}\rvec{x}_{k}-\vec{c}}^2.
\end{align}

Then, we define the channel estimation problem as 
\begin{align}
     \arg\min\limits_{
     \Rbhat_{k}} \norm{\b{X}_k-\Rbhat_k\Hbhat_{k}\Sbhat_k}^2 \quad\text{subject to} \quad 
     {\Rbhat_k}{\Rbhat_k}^\dag  = \I_2\label{eq:ULS_Opt_Problem},
\end{align}
which is known as the orthogonal procrustes problem. The optimal solution is given in \cite{Gibson_ULS_Proof:1962,Schonemann_ULS_Proof:1966,Kabsch:1976}  as
\begin{align}
    	\Rbhat_k = \b{U}_k\b{V}_k^\dag,\label{eq:ULS_Opt_Solution}
\end{align}
where 
\begin{align}\label{eq:Kabsch_alg}
    \b{U}_k\Sigma_k \b{V}_k^\dag = \text{svd}(\b{X}_k\Sbhat_{k}^\dag\Hbhat_k ^\dag),
\end{align}
where svd($\cdot$) stands for singular value decomposition and $\Sigma_k$ is a positive definite diagonal matrix (not used in the algorithm) containing the singular values. Note that $\Rbhat_k$ is not an estimate of $\Rb_k$. Instead it is an averaged estimation of all the rotations in one equalization window. Finally, initializing the estimated channel matrix as $\Hbhat_0 = \I_2$, the next estimated channel will be updated by
\begin{align}
    \Hbhat_{k+\nu} = \Rbhat_k\Hbhat_{k},
\end{align}
where $\nu$ is the sliding stride of the equalization window, meaning that the equalization window slides $\nu$ symbols at a time. The complexity of the algorithm is roughly $L/\nu$ times higher than the \ac{DD-Kabsch} algorithm in \cite{Louchet:2014}.
\begin{figure}
    \centering
    \includegraphics[scale=.4]{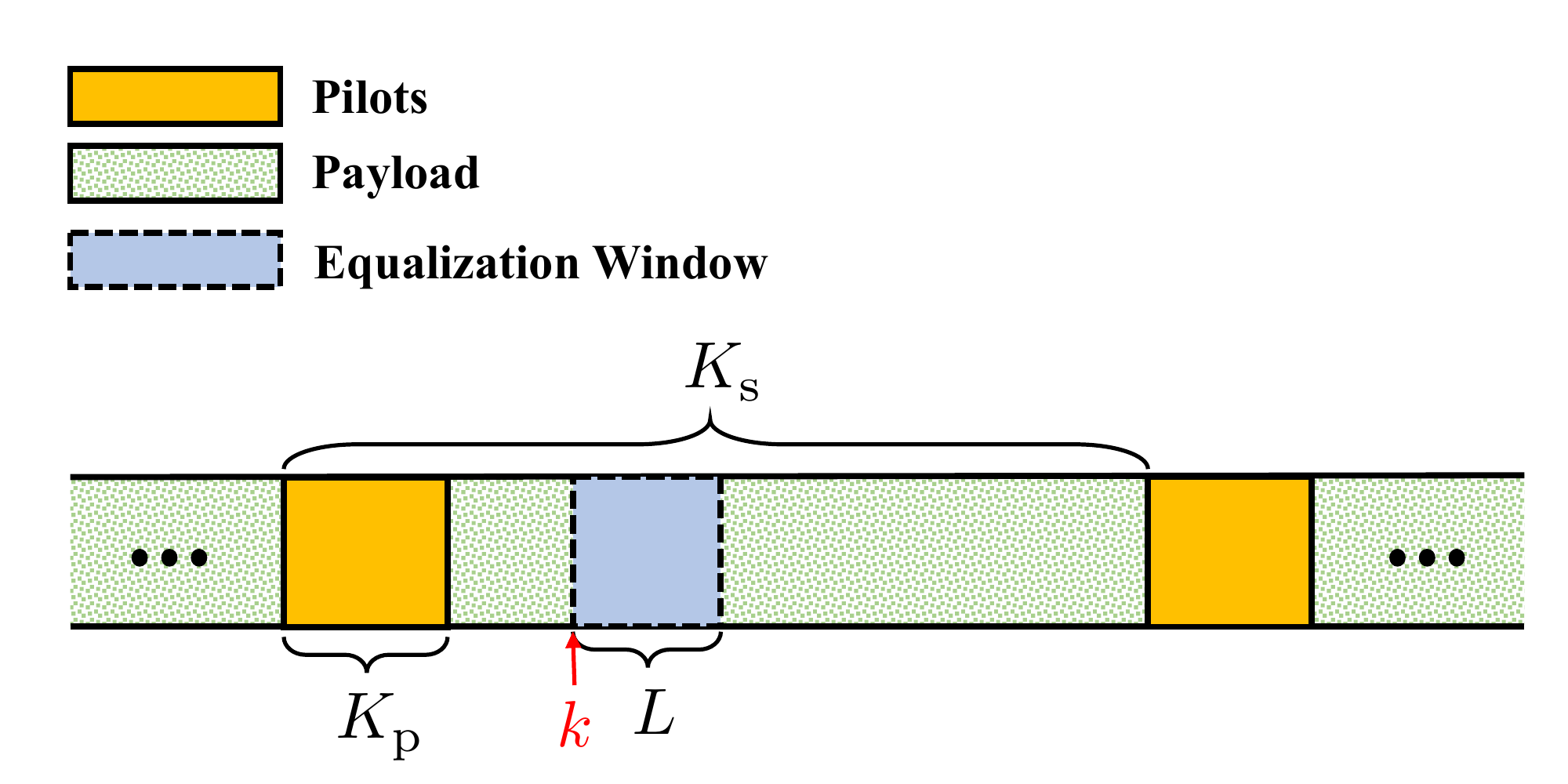}
    \caption{Data frame with inserted pilot sequence and pilot symbols. The transmission block length is $\Ks$ and the length of the pilot sequence is $\Kp$.}
    \label{fig:pilotDataFrame}
\end{figure}
However, we will show that, unlike the \ac{GD}-based algorithms, once $L$ and $\nu$ are set properly, no further active parameter (e.g., step-size) adjustment is needed. 

\begin{algorithm}
        \caption{SW-Kabsch}
        \textbf{Input:} $\b{X}_k$, $\Hbhat_k$, $k$, $\nu$\newline
         \textbf{Output:} $\Hbhat_{k+\nu},\tilde{\Sb}_k, k$
        \begin{algorithmic}[1]
        \STATE\For{$l = 0,\dots,L-1$}{
        \bindent
         \STATE$ \hat{\rvec{s}}_{k+l} = \arg\min\limits_{\vec{c}\in\mathcal{S}}\norm{\Hbhat^{-1}_{k}\rvec{x}_{k+l}-\vec{c}}^2$  \textcolor{black}{// Eq. \eqref{eq:dd_update}} 
         \eindent
        }\vspace{0.1cm}
        \STATE Compute $\b{U}_k$ and $\b{V}_k$ using Eq.~\eqref{eq:Kabsch_alg}
        \STATE$\Rbhat_k =\b{U}_k\b{V}_k^\dag$
        \STATE $ \Hbhat_{k+\nu} = \Rbhat_k\Hbhat_{k}$ \vspace{0.1cm}
        \STATE $\Tilde{\mathbf{S}}_k = [\rvec{\hat{s}}_k,\rvec{\hat{s}}_{k+1},\dots,\rvec{\hat{s}}_{k+\nu-1}]$
        \STATE $k = k+\nu$
    \end{algorithmic}\label{alg:SW-Kabsch}
\end{algorithm}
Algorithm \ref{alg:SW-Kabsch} describes the proposed \ac{SW-Kabsch} algorithm. It takes as inputs the previous instance of the estimated channel $\Hbhat_k$, the $2\times L$ received symbols matrix $\mathbf{X}_{k}$, time instance $k$, and sliding stride $\nu$. Then, it returns the updated channel matrix $\Hbhat_{k+\nu}$, the $2\times \nu$ matrix of decided symbols $\Tilde{\Sb}_k$, and the updated time instance $k$.
\subsection{Pilot-Aided Algorithms}
In this part, we propose a pilot-aided adaptive algorithm to estimate and track both \ac{SOP} drift and accumulated \ac{PDL}. The inserted pilots are chosen from the \ac{QPSK} constellation. 

Fig. \ref{fig:pilotDataFrame} shows the data frame construction where a pilot sequence of length $\Kp$ is inserted at the beginning of each transmission block of length $\Ks$ and as previously defined, $L$ is the length of the equalization window. We also define the $2\times \Kp$ matrix of pilots as
\begin{align}\label{eq:def_pilot}
  \Sp = [\sp_{0},\sp_{1},\dots,\sp_{\Kp -1}],
\end{align}
where the pilots are orthogonal with respect to each polarization (i.e., $\mathrm{S}^\mathrm{p} {\mathrm{S}^\mathrm{p}}^\dag = \delta\I_2$ where $\delta \in \mathbb{R}$ is a constant). The pilot sequence length is set to be larger than $2$, which coincides with the optimal pilot selection suggested in \cite{marzetta_opt_pilot:1999,Hassibi:2003}.
By tuning $\Kp$ and $\Ks$, the performance of the proposed data-aided algorithm can be optimized for different channels; however, there is a trade-off between the overhead and the performance. 

\subsubsection{Sliding Window Least Square Algorithm (SW-LS)}\label{subSec:SW-LS}
The key idea is similar to \ac{SW-Kabsch}. \label{cmr1:wiener_part2}\hlightRI{Note that the standard \ac{LS} algorithm is designed to estimate deterministic channels; however, we combine LS with a decision-directed sliding window enabling us to track the \ac{DP-PDL} channel adaptively.} Considering the \ac{PDL} and \ac{SOP} drift, we can write the next instance of the channel matrix as
\begin{align}\label{eq:SW_LS_Update}
      \Hb_{k+1} = \Gb_k\Hb_{k},
\end{align}
where $\Gb_k$ is $2\times 2$ a complex matrix accounting for the overall rotations and aggregated \ac{PDL} \label{cmr2:cm3_since}\hlightRII{since} time instance $k$. The channel estimation problem can be defined as 
\begin{align}
     \arg\min\limits_{
     \Gbhat_k} \norm{\b{X}_k-\Gbhat_k\Hbhat_{k}\Sbhat_k}^2 \label{eq:PDL_Problem},
\end{align}
where the decision-directed transmitted symbols can be estimated by \eqref{eq:dd_update}. Hence, $\Sbhat_k$ and $\b{X}_k$ can be formed in the same way as \eqref{eq:rxSym_Def}. The optimal solution of \eqref{eq:PDL_Problem} is given by the \ac{LS} algorithm as
\begin{align}\label{eq:AdaptLS}
    \Gbhat_k = \b{X}_k\Sbhat_{k}^\dag\Hbhat_k ^\dag \left(\Hbhat_k \Sbhat_{k}\Sbhat_{k}^\dag\Hbhat_k^{\dag} \right)^{-1}.
\end{align}
Initializing the estimated channel matrix as $\Hbhat_0 = \I_2$, the estimated channel $\Hbhat_{k}$ then is updated by
\begin{align}
    \Hbhat_{k+\nu} = \Gbhat_k\Hbhat_{k}.
\end{align}
Since $\Sbhat_k$ is not hand-picked and comes from the received data symbols, the matrix inverse at the right-hand side of the \eqref{eq:AdaptLS} might become singular. A way to deal with this problem is to use the \ac{MMSE} criterion, which regularize the inverse matrix using he covariance matrix of the \ac{ASE} noise. However, we assume that the receiver has no knowledge of the \ac{ASE} noise power, and hence we do not consider the \ac{MMSE} estimator. Besides, we observed that for high order modulations ($ M\ge 16$) and $L\ge 16$, the chance of running to singularity is low. 

The complexity of the proposed algorithm is approximately $L/\nu$ times higher than \ac{LS} and can be controlled by proper selection of $\nu$ and $L$. 

Algorithm \ref{alg:SW-LS} details the pilot-aided \ac{SW-LS} algorithm. It takes as inputs the received symbols $\mathbf{X}_k$ defined in \eqref{eq:rxSym_Def}, the previous estimate of the channel $\Hbhat_k$, time index $k$, and sliding stride $\nu$ as inputs, and returns the updated channel matrix $\Hbhat_{k+\nu}$, the $2\times \nu$ matrix of decided symbols $\Tilde{\Sb}_k$, and the updated time instance $k$. 
\begin{algorithm}
        \caption{Pilot-Aided SW-LS}
        \textbf{Input:} $\b{X}_k$, $\Hbhat_k$, $k$, $\nu$\\
        \textbf{Output:} $\Hbhat_{k+\nu},\tilde{\Sb}_k, k$
        \begin{algorithmic}[1]
        \STATE\For{$l = 0,\dots,L-1$}{
         \bindent
         \STATE {$i = (k+l\mod \Ks)$}
         \STATE \textbf{if} $i\le \Kp-1$ \textbf{then} \textcolor{black}{ // Check if the symbol is pilot} 
         \INDSTATE $\hat{\rvec{s}}_{k+l} = \sp_{i}$
         \STATE \textbf{else}
         \INDSTATE{$\hat{\rvec{s}}_{k+l} = \arg\min\limits_{\vec{c}\in\mathcal{S}}\norm{\Hbhat^{-1}_{k}\rvec{x}_{k+l}-\vec{c}}^2$  {\textcolor{black}{// Eq. \eqref{eq:dd_update}}} 
        }
        \eindent
        }
        \STATE   $\Gbhat_k  = \b{X}_k\Sbhat_{k}^\dag\Hbhat_k ^\dag \left(\Hbhat_k \Sbhat_{k}\Sbhat_{k}^\dag\Hbhat_k^{\dag} \right)^{-1}$ \textcolor{black}{// Eq.~\eqref{eq:AdaptLS} \vspace{0.1cm}}
        \STATE $ \Hbhat_{k+\nu} = \Gbhat_k\Hbhat_{k}$
        \STATE $\Tilde{\mathbf{S}}_k = [\rvec{\hat{s}}_k,\rvec{\hat{s}}_{k+1},\dots,\rvec{\hat{s}}_{k+\nu-1}]$
        \STATE $k = k+\nu$
        \vspace{0.1cm}
    \end{algorithmic}\label{alg:SW-LS}
\end{algorithm}
\subsubsection{Pilot-Aided Hybrid Algorithms}\label{subSec:hybridAlg}
\label{cmr2:hybrid_alg} \hlightRII{We consider a \ac{DP-PDL} channel with constant \ac{PDL} components $\Gamma_n$. We assume that} for a short enough transmission block length $\Ks$, \hlightRII{the position of the maximum singular value of the channel matrix $\Hb_k$ remains the same, and that} the aggregated \ac{PDL} ratio $\boldsymbol{\rho}_k$ is almost constant.
Thus, one can argue that after compensating for the channel using the pilots, the residual channel matrix is almost unitary, which can be tracked using an arbitrary unitary channel tracking algorithm. 
Therefore, considering the frame structure in Fig. \ref{fig:pilotDataFrame}, we propose two hybrid algorithms, which are described in the following.

Since the frame structure is periodic, the algorithms are described for the first frame, where the same procedure is repeated for the subsequent frames. The first $\Kp$ transmitted symbols are pilots as defined in \eqref{eq:def_pilot} and the corresponding $2\times \Kp$ matrix of received symbols is $\mathbf{X}^\mathrm{p} = [\rvec{x}_0,\dots,\rvec{x}_{\Kp -1}]$. We define the set of payload indices as $\mathcal{P} = \{\Kp+1,\dots,\Ks\}$. In the first stage of the hybrid algorithms, a coarse estimation of the channel is obtained by applying the \ac{LS} estimator as
\begin{align}
    \tilde{\Hb}_{0} = \frac{1}{\delta}\mathbf{X}^\mathrm{p}\Sp.
\end{align}
The compensated payload symbols can be obtained according to
\begin{align}\label{eq:hybrid_comp}
    \mathbf{X}_\text{c} =  \tilde{\Hb}_{0}^{-1} \mathbf{X}^\mathrm{d},
\end{align}
where $\mathbf{X}^\mathrm{d} = [\rvec{x}_{\Kp},\dots, \rvec{x}_{\Kp+\Ks-1}]$ is the matrix of payload symbols. The residual channel matrix at time $k \in \mathcal{P}$ can be written as
\begin{align}\label{eq:residual_chan}
    \Hbr_{k} = \tilde{\Hb}_{0}^{-1}\mathbf{H}_{k}.
\end{align}
For a short $\Ks$, the aggregated \ac{PDL} ratio $\boldsymbol{\rho}_k$ for $k \in \mathcal{P}$ is assumed to be constant. Thus, assuming that the aggregated \ac{PDL} is compensated by $\tilde{\Hb}_{0}^{-1}$, the residual channel matrix $\Hbr_{k}$ can be regarded as an almost unitary matrix. Therefore, in the second stage, given $\mathbf{X}_\text{c}$, a unitary adaptive algorithm is used to track the residual channel matrix $\Hbr_{k}$ for $k\in\mathcal{P}$. 

In this paper, using \ac{LS} for the first stage, we propose the \ac{LS}-\ac{SW-Kabsch} and \ac{LS}-\ac{DD-Czegledi} hybrid algorithms where \ac{SW-Kabsch} and \ac{DD-Czegledi} \cite{Czegledi:2016} are used in the second stage, respectively.
\section{Results}\label{sec:Results}
\subsection{Simulation Setup}
\Ac{PM-16-QAM} at a symbol rate of $\smash{R_s = 28}$ Gbaud (i.e., $T = 1/Rs_{s}$) is considered. A random sequence of $\smash{K = 10^5}$ symbols is transmitted on each polarization where the initial matrices $\Jb_{0,1},\dots, \Jb_{0,N}$ are drawn from the set of all $2\times 2$ unitary matrices. Thereby the \ac{SOP} gets a uniform distribution over the Poincar\'e sphere. The presented results are evaluated by averaging over $10^{5}$ such sequences. For the \ac{DP-PDL} channel, it is assumed that the link has $\smash{N = 20}$ segments where all the segments have identical polarization linewidth $\Deltap$ and segment-wise \ac{PDL} ratio $\gamma_{n}$. The launch power in each polarization is $P = \E[\rvec{s}_k^\dag \rvec{s}_k]$ and the \ac{SNR} per polarization is defined as $\text{SNR} = P/\sigma^{2}_{z}$. 
The performance is assessed by estimating the \ac{SER} for different setups in the presence of \ac{SOP} drift, \ac{PDL}, and \ac{ASE} noise.

\subsubsection{Blind~/~Decision-Directed Algorithms Setup}
For comparison, the results obtained from the \ac{MCMA}\cite{oh_MCMA:1995}, \ac{DD-Kabsch} \cite{Louchet:2014}, \ac{DD-Czegledi} \cite{Czegledi:2016} algorithms are presented as benchmarks. The \ac{DD-Kabsch} algorithm operates in a decision-directed block-wise fashion where it uses a block size of $L_\text{Kabsch} = 16$, which gives the best polarization drift tolerance for this algorithm. To ensure the convergence of the \ac{DD-Czegledi} and \ac{MCMA} algorithms, they are implemented with two stages of convergence where the first stage uses a larger tracking step size $\mu$ than the one is used in the second stage; for detailed implementation, refer to \cite[Algorithm 1]{Czegledi:2016}. To compare the \ac{SW-Kabsch} algorithm with the best version of the benchmarks, the tracking step size $\mu$ of the \ac{DD-Czegledi} and \ac{MCMA} algorithms is optimized as a function of \ac{SOP} drift speed and \ac{SNR}. 

Moreover, coherent differential coding is used to resolve the four-fold phase ambiguity of the \ac{PM-16-QAM} constellation for blind and decision-directed algorithms. \label{cmr2:differential}\hlightRII{Deploying differential coding induces an \ac{SNR} penalty meaning that an extra \ac{SNR} is needed to attain a given \ac{SER} compared to nondifferential schemes \cite{weber_DiffEncoding:1978}.} For comparison, differential coding is used even when the channel is perfectly known (i.e., $\Hbhat_k = \Hb_k$). Both decision-directed and blind algorithms may swap the equalized channels yielding polarization ambiguity. The polarization ambiguity may be resolved by inserting a few pilots in the data load. In most communication systems, it is common to use pilots for different purposes such as timing synchronization, carrier frequency estimation, etc. Therefore, in this paper, a genie-aided ambiguity resolution is used for both benchmarks and the proposed algorithm in our simulation, whereas we expect real deployed receivers to use pilots.

The \ac{SW-Kabsch} operates with a sliding window of size $\smash{L = 24}$. The sliding stride is $\smash{\nu = 6}$, which results in $4$ times higher computational complexity than the \ac{DD-Kabsch} algorithm. The parameters $L$ and $\nu$ are chosen such that \ac{SW-Kabsch} outperform the benchmarks with relatively low complexity. Note that optimizing $\nu$ and $L$ as a function of \ac{SOP} drift might alter the performance of the algorithm for a fixed computational complexity. 
\begin{figure}
    \centering
     \includegraphics{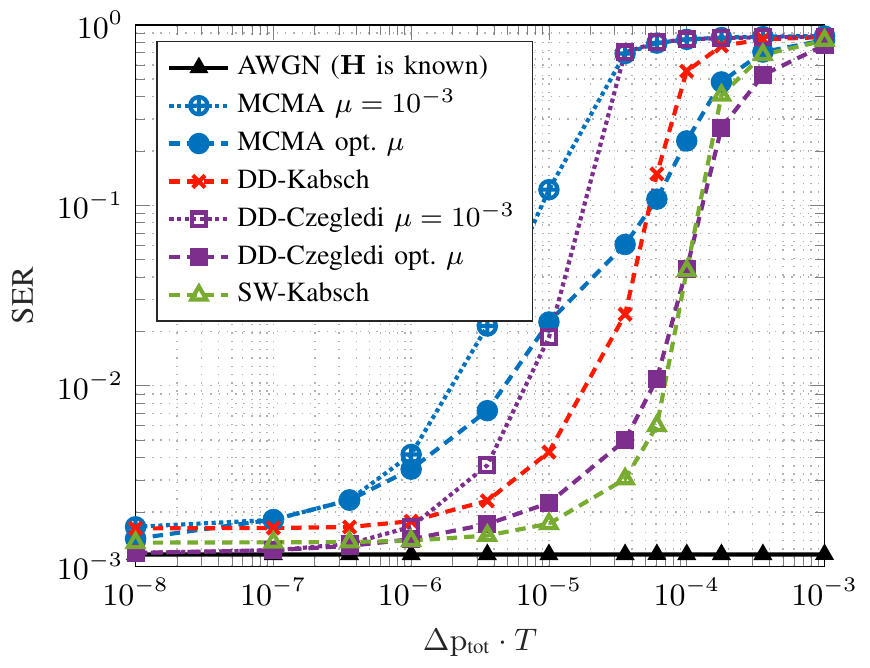}
    \caption{The polarization drift tolerance at $\smash{\text{\ac{SNR}}=18}$ dB is shown for four tracking algorithms using \ac{PM-16-QAM} constellations \label{cmr2:SER}\hlightRII{where the vertical axis shows the symbol error rate (SER)}.}
    \label{fig:SER_vs_SOP}
\end{figure}
\begin{figure*}
\centering
 \includegraphics{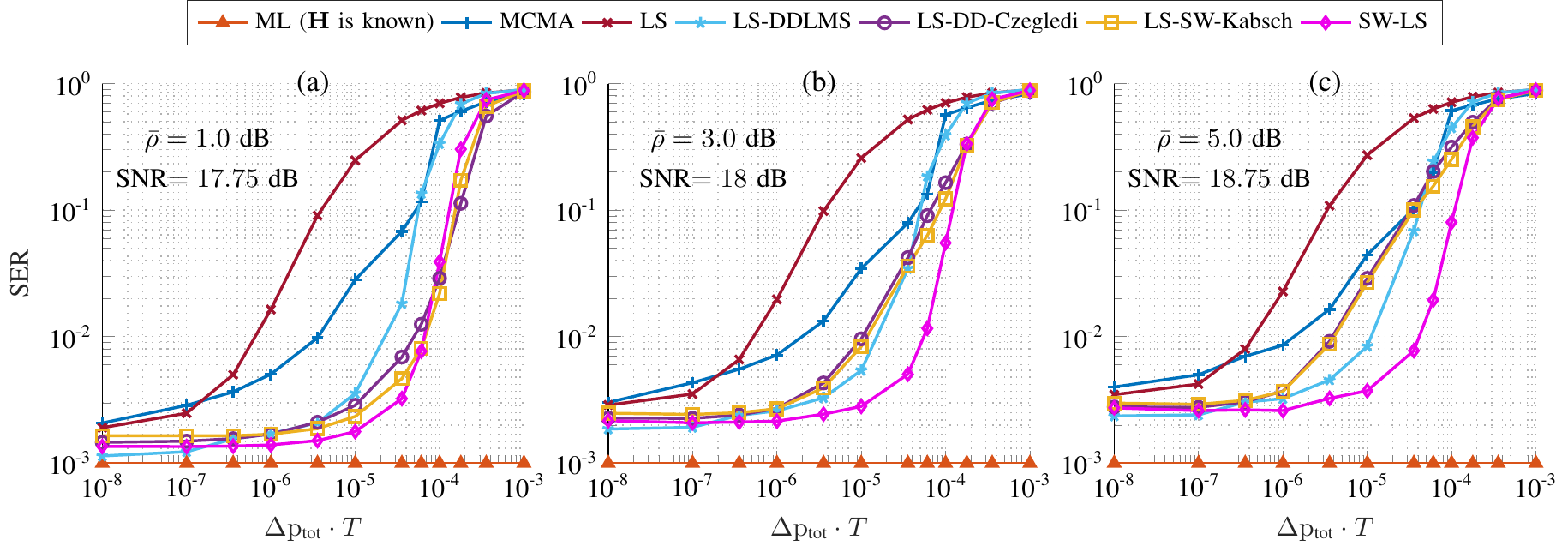}
\caption{\label{fig:PDL_SER_vs_Deltap} The achievable \ac{SER} performance of pilot-aided algorithms for a fiber with $\smash{N = 20}$ segments is plotted. In (a), (b), and (c), to change the average aggregated \ac{PDL} $\boldsymbol{\rho}$, the segment-wise \ac{PDL} ${\varphi}_n$ of all the segments is set to $0.25$, $0.70$, and $1.10$ dB, respectively.}
\end{figure*}
\subsubsection{Pilot-aided Algorithms Setup}
The pilot length is chosen to be $\smash{\Kp = 16}$, and the block length is assumed to be $\smash{\Ks = 1016}$ symbols yielding $1.6\%$ overhead.
Note that optimizing $\Kp$ and $\Ks$ as a function of the polarization linewidth, $\bar{\rho}$, and \ac{SNR} might improve the performance of the pilot-aided tracking algorithms; however, this requires a feedback channel between the receiver and the transmitter, which might not be available in many applications. Thus, we used fixed $\Kp$ and $\Ks$, which performs reasonably well for a large range of channel parameters.
The \ac{SW-LS} algorithm is set to use an equalization window of $\smash{L=24}$ with a $\nu=6$ to get a fair performance while keeping the complexity low.  

As for the benchmarks, a hybrid algorithm called \ac{LS}-\ac{DDLMS} is used, where the \ac{DDLMS} algorithm is initialized with a coarse estimate of the channel obtained by applying \ac{LS} on the pilot sequence. The \ac{ML} detection for a known channel at the receiver is also used to serve as a benchmark.

\subsection{Polarization Drift Tolerance}
This part evaluates the channel tracking ability of the proposed and conventional algorithms for the \ac{DP-PDL} channel. The polarization drift sensitivity is measured by sweeping $\Deltap$ while the \ac{SNR} is fixed. \label{cmr1:deltap_part2}\hlightRI{The polarization movement between two consecutive symbols on the Poincar\'e sphere can be quantitatively characterized by the angle of the polarization movement $\theta_\mathrm{SOP}$ \cite[Eq.~(2)]{Charlton:2017}. For a fixed symbol rate of $R_s= 28$ Gbaud (i.e., $T = 35.7~\mathrm{ps}$), we sweep $\Deltaptot\cdot T$ from $10^{-8}$ to $10^{-3}$ which gives an average $\theta_\mathrm{SOP}$ from $5\cdot10^{-4}$ to $1.57\cdot10^{-1}$ rad/symbol, respectively. Note that $\theta_\mathrm{SOP}$ does not correspond to the angular velocity of the \ac{SOP} movement in rad/s. The angular velocity is quantified as the time derivative of $\theta_\mathrm{SOP}$, which is undefined for our model (i.e., the time derivative of a Wiener process is undefined).}
\subsubsection{Channel with Negligible \ac{PDL}}
The \ac{SER} versus $\smash{\Deltaptot\cdot T}$ is plotted in Fig. \ref{fig:SER_vs_SOP}. The \ac{SNR} is set to achieve $\smash{\text{\ac{SER}}=10^{-3}}$ for a known channel at the receiver. As can be seen, the proposed \ac{SW-Kabsch} algorithm offers a better polarization drift tolerance than the benchmarks at the expense of approximately $4$ times higher complexity than \ac{DD-Kabsch}. More specifically, when the channel drifts slowly, i.e., $\smash{\Deltaptot \cdot T < 10^{-6}}$, \ac{SW-Kabsch} and \ac{DD-Czegledi} show almost the same performance; \ac{SW-Kabsch} gradually outperforms \ac{DD-Czegledi} when the channel gets faster. Finally, \ac{DD-Czegledi} takes over at $\smash{\Deltaptot \cdot T > 10^{-4}}$, but in this regime, the \ac{SER} is out of practical interest. 

To investigate the step size sensitivity of the \ac{GD}-based algorithms, the \ac{SER} of \ac{MCMA} and \ac{DD-Czegledi} for fixed and optimal $\mu$ is also plotted (see blue and purple dotted curves). The step size is set to $\mu = 10^{-3}$ such that \ac{MCMA} and \ac{DD-Czegledi} perform their best for slowly varying channels. Evidently, \ac{MCMA} and \ac{DD-Czegledi} show a higher step size sensitivity and hence a lower polarization drift tolerance for a fixed $\mu$.
While the tracking capability of the \ac{GD}-based algorithms is highly dependent on the proper adjustment of the tracking step size, \ac{SW-Kabsch} requires no parameter tuning. This could be advantageous in bursty channels where the \ac{SOP} drift speed does not remain constant during the whole transmission. Thus, an algorithm that is tailor-made for a specific \ac{SOP} drift speed may fail to track sudden changes in the channel. 

\begin{figure*}
\centering
 \includegraphics{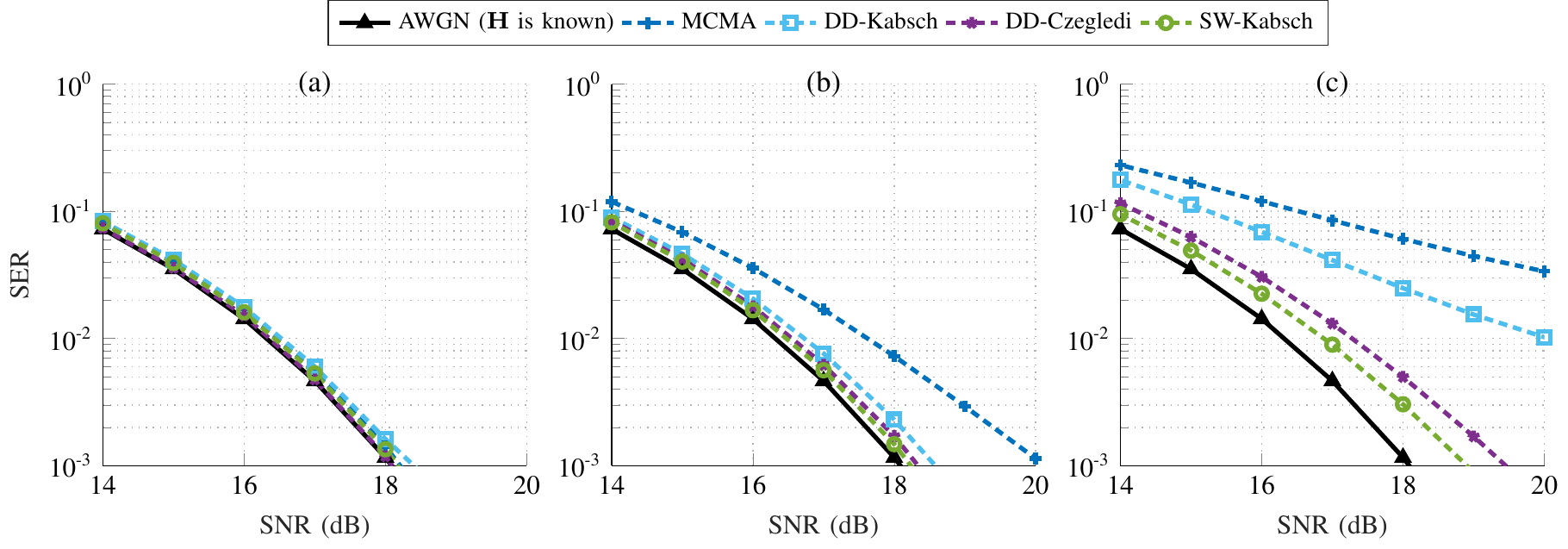}
\caption{\label{fig:SOP_SER_vs_SNR} The achievable \ac{SER} performance of the blind tracking algorithms when \ac{PDL} is negligible, for three drift speeds is depicted where ({a}) $\smash{\Deltaptot\cdot T = 10^{-8}}$, ({b}) $\smash{\Deltaptot\cdot T = 3.57\cdot10^{-6}}$, and ({c}) $\smash{\Deltaptot\cdot T = 3.57\cdot 10^{-5}}$.}
\end{figure*}
\begin{figure*}
\centering
 \includegraphics{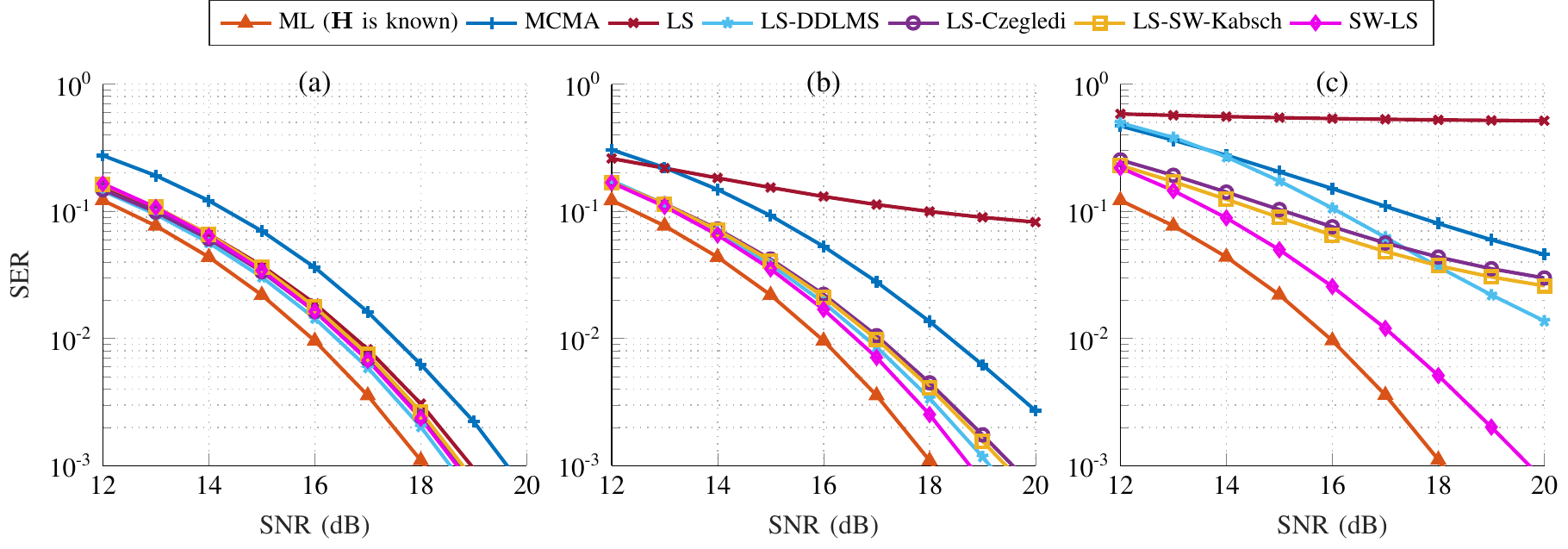}
\caption{\label{fig:PDL_SER_vs_SNR} The achievable \ac{SER} performance of various algorithms for a \ac{DP-PDL} channel with an average aggregated \ac{PDL} of $\bar{\rho}=3.0$ dB is plotted where (\textbf{a}) $\smash{\Deltaptot\cdot T = 10^{-8}}$, (\textbf{b}) $\smash{\Deltaptot\cdot T = 3.57\cdot10^{-6}}$, and (\textbf{c}) $\smash{\Deltaptot\cdot T = 3.57\cdot 10^{-5}}$.}
\end{figure*}
\subsubsection{Channel with considerable \ac{PDL}}
Fig. \ref{fig:PDL_SER_vs_Deltap} shows the \ac{SER} versus $\smash{\Deltaptot\cdot T}$ for three different average aggregated \ac{PDL} $\bar{\rho}$ levels where the \ac{SNR} is set such that $\smash{\text{\ac{SER}}=10^{-3}}$ is achieved with the \ac{ML} detection. Note that to change the average aggregated \ac{PDL} ${\bar{\rho}}$ in Figs. \ref{fig:PDL_SER_vs_Deltap}(a), (b), and (c), the segment-wise \ac{PDL} ${\varphi}_n$ \label{varphi}\hlightRII{(defined in \eqref{eq:varphi})} of all the segments is set to $0.25$, $0.70$, and $1.10$ dB, respectively. 
It can be seen that \ac{SW-LS} shows the best tolerance to \ac{SOP} drift in all considered average aggregated \ac{PDL} $\bar{\rho}$ levels. All pilot-aided algorithms behave roughly the same at low \ac{SOP} drifts $\smash{(\Deltaptot\cdot T < 10^{-6})}$, at higher drift speeds $\smash{(\Deltaptot\cdot T \ge 10^{-6})}$, \ac{SW-LS} shows the best polarization drift tolerance.

Interestingly, for $\smash{\bar{\rho} =1.0~\text{dB}}$, the hybrid unitary tracking algorithms (\ac{LS}-\ac{DD-Czegledi} and \ac{LS}-\ac{SW-Kabsch}) have a better polarization drift tolerance than the \ac{LS}-\ac{DDLMS} algorithm. A possible justification might be that for small $\bar{\rho}$, the residual channel $\Hb^{r}_{k}$ \eqref{eq:residual_chan} is almost unitary, and therefore a unitary estimation of the channel might result in a better estimate. This implies that even for $\smash{\bar{\rho}=1.0~\text{dB}}$, nonunitary tracking algorithms could be replaced by unitary tracking algorithms to obtain higher polarization drift tolerance. The performance of the proposed hybrid algorithms might alter for larger $\bar{\rho}$ by decreasing the transmission block length $\Ks$ at the expense of higher overhead.

\subsection{Additive Noise Tolerance}
Fig.~\ref{fig:SOP_SER_vs_SNR} compares the \ac{SER} of \ac{SW-Kabsch} with the benchmarks for a \ac{DP-PDL} channel for three different \ac{SOP} drift speeds. The \ac{PDL} is assumed to be negligible (i.e., $\smash{\gamma_n = 0}$). All the studied algorithms have roughly similar performances for a very slowly drifting channel, see Fig.~\ref{fig:SOP_SER_vs_SNR}(a).
However, when the channel drifts quickly, as shown in Fig.~\ref{fig:SOP_SER_vs_SNR}(b), \ac{DD-Czegledi} and \ac{SW-Kabsch} perform almost similarly, and \ac{MCMA} is no longer an option. Finally, for an even faster channel presented in Fig.~\ref{fig:SOP_SER_vs_SNR}(c), \ac{SW-Kabsch} outperforms all the other algorithms.

Fig.~\ref{fig:PDL_SER_vs_SNR} shows the \ac{SER} of various algorithms for a \ac{DP-PDL} channel with $\smash{\bar{\rho} = 3.0~\text{dB}}$. From Fig. \ref{fig:PDL_SER_vs_SNR}(a), it can be concluded that all the considered algorithms except \ac{MCMA} perform similarly for a slowly drifting channel. However, for fast drifting channels, Figs. \ref{fig:PDL_SER_vs_SNR}(b) and \ref{fig:PDL_SER_vs_SNR}(c) show that the proposed \ac{SW-LS} algorithm outperforms the other algorithms. 
\subsection{Computational Complexity vs. Performance}
In \cite[Table I]{Czegledi:2016}, it has been shown that \ac{DD-Czegledi} is almost twice as complex as \ac{DD-Kabsch} for a dual-polarization channel. Therefore, taking \ac{DD-Kabsch} as a reference, a rough complexity comparison of the proposed algorithm is presented. 
Compared to \ac{DD-Kabsch}, the \ac{SW-Kabsch} algorithm has an additional sliding window, resulting in approximately $L/\nu$ times higher complexity. The \ac{SW-LS} algorithm has a matrix inverse instead of the singular value decomposition in \ac{SW-Kabsch}, which essentially has the same complexity. Therefore, \ac{SW-LS} is also at least $L/\nu$ times more complex than \ac{DD-Kabsch}. \label{cmr2:complexity}\hlightRII{While the complexity of the proposed algorithms is tractable for low-dimensional channels (e.g., \ac{DP-PDL} channel, few-mode fiber, etc.), it becomes challenging for high-dimensional channels} since the complexity of singular value decomposition and matrix inversion operations increases \label{author:complexity}\highlight{cubically} with the number of dimensions.

The complexity of the algorithms can be decreased \label{cmr2:performance}\hlightRII{at the expense of performance}. For instance, reducing the $L/\nu$ ratio decreases the complexity of the proposed algorithms but also degrades the performance. Besides, adjusting the $L/\nu$ ratio as a function of channel parameters might reduce the complexity of the algorithms for certain channel conditions (e.g., slowly drifting channels).
\section{Conclusion}
We have proposed \ac{SW-Kabsch} and \ac{SW-LS} algorithms to track memoryless \ac{DP-PDL} channels. Both proposed algorithms use a sliding window in a decision-directed fashion, processing multiple symbols at a time. Numerical simulations are used to evaluate and compare the performance of the proposed algorithms with the conventional tracking algorithms, including \ac{MCMA} which is the most popular tracking algorithm in the literature. Results show that the proposed algorithms are more robust to polarization drift than the benchmarks. While parameter adjustment is required for \ac{GD}-based benchmarks (e.g., \ac{MCMA}, \ac{DDLMS}, \ac{DD-Czegledi}, etc.), the proposed algorithms need no such adjustments and still outperform the benchmarks. Besides, unlike the \ac{DD-Czegledi} algorithm \cite{Czegledi:2016}, which cannot be scaled for higher dimensional channels, the proposed algorithm can be applied to any number of dimensions at the expense of increasing the computational complexity. 

The proposed algorithms are analyzed assuming negligible polarization-mode dispersion, which is not the case in more realistic fiber channels. \label{cmr2:SOP_comp}\hlightRII{Although it is not possible to compensate for polarization-mode dispersion and \ac{SOP} separately, the proposed algorithms can be used in a hybrid fashion as in \cite{zheng_PMD_SOP:2018}, where polarization-mode dispersion is compensated in the frequency domain and \ac{SOP} is tracked in the time domain.}
The effect of polarization-mode dispersion is left for future work.

\bibliography{references}
\vfill
\end{document}